\documentclass[a4paper]{spie}
\usepackage{graphicx}
\title{Proposal of fault-tolerant tomographic image reconstruction}
\author{Hiroyuki Kudo\supit{a,b}, Keita Takaki\supit{a}, Fukashi Yamazaki\supit{a}, and Takuya Nemoto\supit{a}
\skiplinehalf
\supit{a}Faculty of Engineering, Information and Systems, University of Tsukuba, Tennoudai 1-1-1, Tsukuba 305-8573, Japan \\
\supit{b}JST-ERATO Momose Quantum-Beam Phase Imaging Project, Katahira, Aoba-ku, Sendai 980-8577, Japan
}
\authorinfo{Further author information: (Send correspondence to H.K.)\\H.K.: E-mail: kudo@cs.tsukuba.ac.jp}
\begin{document}
\maketitle
\begin{abstract}
This paper deals with tomographic image reconstruction under the situation where some of projection data bins are contaminated with abnormal data. Such situations occur in various instances of tomography. We propose a new reconstruction algorithm called the Fault-Tolerant reconstruction outlined as follows. The least-squares ($L^2$-norm) error function $\parallel A\vec{x}-\vec{b} \parallel_2^2$ used in ordinary iterative reconstructions is sensitive to the existence of abnormal data. The proposed algorithm utilizes the $L^1$-norm error function $\parallel A\vec{x}-\vec{b} \parallel_1^1$ instead of the $L^2$-norm, and we develop a row-action-type iterative algorithm using the proximal splitting framework in convex optimization fields. We also propose an improved version of the $L^1$-norm reconstruction called the $L^1$-TV reconstruction, in which a weak Total Variation (TV) penalty is added to the cost function. Simulation results demonstrate that reconstructed images with the $L^2$-norm were severely damaged by the effect of abnormal bins, whereas images with the $L^1$-norm and $L^1$-TV reconstructions were robust to the existence of abnormal bins.
\end{abstract}
\section{INTRODUCTION}
\par This paper deals with tomographic image reconstruction under the situation where some of projection data (sinogram) bins are contaminated with abnormal data, {\it i.e.} impulsive noise with heavy tails, and locations of the abnormal bins are unknown. We believe that such situations occur in various instances of tomography due to unexpected troubles in detector, x-ray source or power supply, data transfer errors, metal in the object, misalignments in scanner geometry, and phase wrapping in phase tomography. However, according to our knowledge, there exist very few (almost no, to the best of our knowledge) literatures which deal with image reconstruction to handle such situations. On the other hand, in other fields such as statistics, computer vision, and machine learning, the problem of estimating some physical parameters from a set of measured data containing the abnormal errors (outliers) has been recognized as an important research topic so that there exist a lot of research activities.
\par We propose a new image reconstruction algorithm called the Fault-Tolerant reconstruction outlined as follows. In ordinary iterative reconstructions, the least-squares error $\parallel A\vec{x}-\vec{b} \parallel_2^2$ defined as the $L^2$ norm difference between forward-projected projection data $A\vec{x}$ and measured projection data $\vec{b}$ is used as a cost function [1],[2]. However, it is well-known that the least-squares criterion is sensitive to the existence of abnormal data in the measurement $\vec{b}$. A simple method to estimate the locations of abnormal bins and exclude the corresponding data from the data fitting is to utilize the $L^1$-norm difference $\parallel A\vec{x}-\vec{b} \parallel_1^1$ between $A\vec{x}$ and $\vec{b}$ instead of the $L^2$ norm [3],[4]. Based on this idea, we develop a row-action-type iterative algorithm minimizing the $L^1$ norm error function using the proximal splitting framework in convex optimization fields [5],[6]. The resulting algorithm resembles the structure of ART (Algebraic Reconstruction Technique) algorithm [7],[8]. We also propose an improved version of the $L^1$-norm approach called the $L^1$-TV reconstruction, in which a weak Total Variation (TV) penalty function is added to the cost function [9]-[11]. We performed simulation studies for typical scenarios assuming detector errors, x-ray radiation errors, and random errors in projection data space. In all of the simulated scenarios, reconstructed images with the $L^2$-norm were severely damaged by the effect of abnormal bins, whereas images with the $L^1$-norm and $L^1$-TV reconstructions were surprisingly robust to the existence of abnormal bins.
\section{PROPOSED METHOD}
\subsection{Abnormal Error Noise Model}
\par This section describes the setup of image reconstruction problem dealt in this paper, and derives the corresponding noise model together with the cost function used in image reconstruction. We denote an image by a $J$-dimensional vector $\vec{x}=(x_1,x_2,\dots,x_J)^T$ and denote a corresponding projection data (sinogram) by an $I$-dimensional vector $\vec{b}=(b_1,b_2,\dots,b_I)^T$. We denote the system matrix which relates $\vec{x}$ to $\vec{b}$ by $A=\{a_{ij}\}$, where we assume that $I>J$, {\it i.e.} the linear equation $A\vec{x}={\vec{b}}$ is overdetermined. Then, the measurement process of projection data is expressed as
\begin{eqnarray}
\vec{b}=A\vec{x}+\vec{n},
\end{eqnarray}
{\par\noindent}where $\vec{n}=(n_1,n_2,\dots,n_I)^T$ denotes measurement noise. In ordinary iterative reconstructions, the cost function is designed by assuming that the noise component $\vec{n}$ follows independent Gaussian distribution or independent Poisson distribution. On the other hand, in this paper, we consider the Abnormal Error Noise (AEN) model, in which most elements of $\vec{n}$ are zeros but only a small number of elements of $\vec{n}$ take non-zero abnormal error values occurred by various physical reasons described in Section 1. We assume that the abnormal projection data takes a value over the pre-specified interval $[A\vec{x}\mid_i-m_1,A\vec{x}\mid_i+m_2]$ with a uniform probability, where $m_1>0,m_2>0$ are pre-specified constants corresponding to the dynamic range to control error severity. To demonstrate that the abnormal errors in only a small number of projection data bins lead to significant artifacts, in Fig. 1, we show examples of reconstructed images for three typical cases, {\it i.e.} data corresponding to one detector element contains the error, data corresponding to one projection angle contains the error, and randomly selected five projection data bins contain the error. Furthermore, we also assume that locations of the abnormal data bins are unknown. Below, we explain that such noise model can be well represented by using a particular case of generalized Gaussian probability density function. In the generalized Gaussian model, the likelihood function of $\vec{b}$ given $\vec{x}$ is expressed as
\begin{eqnarray}
{\rm Pr}(\vec{b}\mid\vec{x})={{1}\over{Z}}\exp(-\alpha\parallel A\vec{x}-\vec{b}\parallel_p^p),
\end{eqnarray}
{\par\noindent}where $Z$ denotes the partition function, $\alpha$ denotes the (variance) parameter of density function, and $\parallel \vec{r}\parallel_p$ denotes the $L^p$-norm with $0\leq p\leq 2$ as the norm parameter. For unfamiliar readers, the definition of $\parallel \vec{r}\parallel_p$ is expressed as
\begin{eqnarray}
\parallel \vec{r}\parallel_p^p =\cases{\displaystyle \sum_{i=1}^I \mid r_i \mid^p & $p\neq 0$ \cr \displaystyle\lim_{\epsilon\rightarrow +0}\displaystyle\sum_{i=1}^I \mid r_i \mid^{\epsilon} & $p=0$},
\end{eqnarray}
{\par\noindent}where $\vec{r}=(r_1,r_2,\dots,r_I)^T$. It is well-known that the noise model corresponding to $p=2$ is the independent Gaussian model, which is used when the additive noise $\vec{n}$ follows the independent Gaussian distribution with a same variance for all the bins. Also, the noise model corresponding to $p=1$ is the independent Laplacian model, which can handle heavy tail noise like impulsive noise better compared with the Gaussian [12]. The most appropriate noise model for the above AEN model is the case of $p=0$ due to the following reason. In the case of $p=0$ in Eq. (2), $\parallel A\vec{x}-\vec{b}\parallel_p^p$ inside $\exp(\cdot)$ becomes the so-called $L^0$-norm, which represents the number of non-zero elements in the noise vector $\vec{n}=\vec{b}-A\vec{x}$. Therefore, the corresponding likelihood ${\rm Pr}(\vec{b}\mid\vec{x})$ becomes large when $\vec{n}=\vec{b}-A\vec{x}$ has only a small number of non-zero elements, which matches to the assumption of AEN model. The similar discussion on the above property of $L^0$-norm can be found in many literatures of Compressed Sensing (CS) [13],[14]. Furthermore, by taking the negative logarithm of Eq. (2), we obtain the cost function of image reconstruction $f(\vec{x})$ to be minimized corresponding to the AEN model as
\begin{eqnarray}
f(\vec{x})=-{\rm Pr}(\vec{b}\mid\vec{x})=\parallel A\vec{x}-\vec{b} \parallel_0^0+{\rm constant}\ \ (L^0{\rm -}{\rm norm}\ {\rm reconstruction}).
\end{eqnarray}
{\par\noindent}However, as is well-known in the CS fields [13],[14], the $L^0$-norm is a non-convex function in which its exact minimization is of NP-hard complexity. So, by following the standard approximation used in the CS fields, we replace the $L^0$-norm by the $L^1$-norm as
\begin{eqnarray}
f(\vec{x})=\parallel A\vec{x}-\vec{b} \parallel_1^1\ \ (L^1{\rm -}{\rm norm}\ {\rm reconstruction}).
\end{eqnarray}
{\par\noindent}This replacement is popular in the CS fields and is called the convex relaxation in optimization literatures [13],[14]. On the other hand, the standard cost function in image reconstruction fields is the $L^2$-norm expressed as
\begin{eqnarray}
f(\vec{x})=\parallel A\vec{x}-\vec{b} \parallel_2^2\ \ (L^2{\rm -}{\rm norm}\ {\rm reconstruction}).\end{eqnarray}
{\par\noindent}Equation (6) is derived from the Gaussian noise model by putting $p=2$ in Eq. (2), which is quite different from the AEN model. Therefore, in image reconstruction under the AEN model, we expect that there exist significant differences in image quality between Eq. (5) and Eq. (6). Actually, in Section 3, we demonstrate that there exist surprisingly large differences between the $L^1$-norm reconstruction and the $L^2$-norm reconstruction.
\par One more remark is as follows. The above derivation inspires that our method is same as image reconstruction using the CS technique. In the CS fields, the $L^1$-norm is normally used in regularization terms to pick up sparse solutions $\vec{x}$ from a large number of feasible solutions. However, we use the $L^1$-norm for data fidelity (likelihood) terms. Such approaches are frequently used in statistics, computer vision, and data mining to mitigate the effect of outliers in the data fitting, with various different terminologies such as robust statistics, robust estimation, and outlier detection [3],[4],[15],[16]. However, according to our knowledge, there exist very few (almost no) literatures to use it in tomographic image reconstructions.
\begin{figure}
\centerline{\includegraphics[height=8cm,clip]{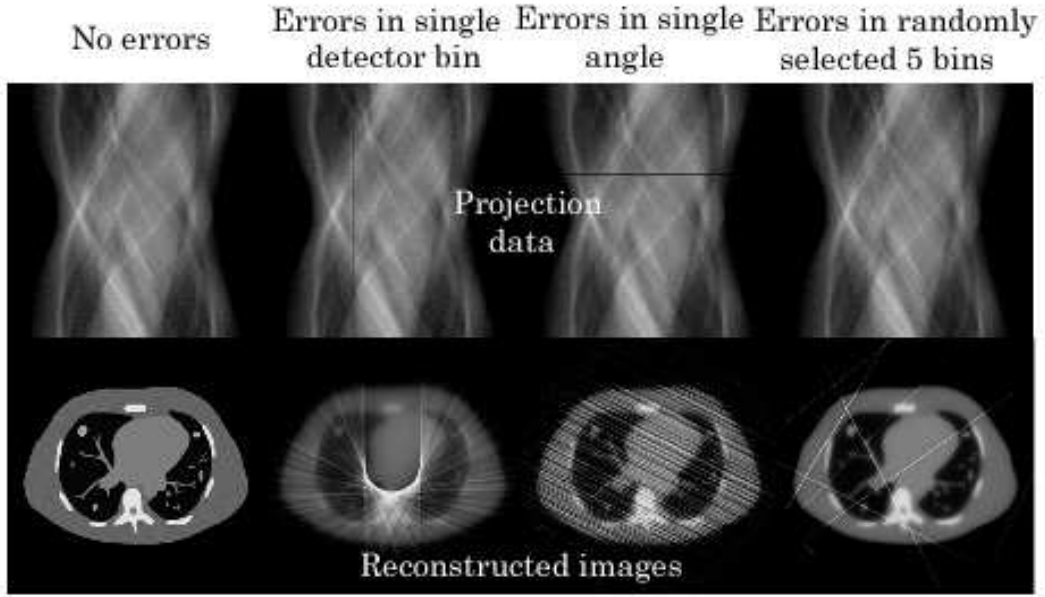}}
\caption{Typical image degradations occurred by the abnormal data bins in projection data. Top images show projection data where black line or dots show the abnormal bins. Bottom images show corresponding reconstructed images.}
\end{figure}
\subsection{Row-Action-Type Iterative Algorithm for $L^1$-Norm Reconstruction}
\par The most significant contribution in this paper is to propose a row-action-type iterative algorithm for the $L^1$-norm image reconstruction, which resembles the structure of famous ART algorithm [7],[8]. The algorithm derivation is based on a framework called the proximal splitting in convex optimization fields [5],[6]. Before going into the concrete derivation, we briefly review the proximal splitting together with some related basic knowledges. Let us consider a convex minimization problem formulated as
\begin{eqnarray}
\min_{\vec{x}} f(\vec{x}),
\end{eqnarray}
{\par\noindent}where we assume that $f(\vec{x})$ is a possibly non-differentiable lower semi-continuous (lsc) convex function.
{\par\noindent}[Proximity Operator and Proximal Algorithm] [17] The proximity (prox) operator corresponding to $f(\vec{x})$ is defined by
\begin{eqnarray}
\vec{x}={\rm prox}_{\alpha f}(\vec{z})\equiv{\rm arg}\min_{\vec{x}} (f(\vec{x})+{{1}\over{2\alpha}}\parallel\vec{x}-\vec{z}\parallel_2^2),
\end{eqnarray}
{\par\noindent}where the parameter $\alpha$ is called the stepsize parameter. Two nice properties of the ${\rm prox}$ operator are that it can be defined for any lsc convex function even if it is non-differentiable like the $L^1$-norm (including convex constraint), and that its fixed points, {\it i.e.} $\vec{x}$ satisfying $\vec{x}={\rm prox}_{\alpha f}(\vec{x})$, coincide with minimizers of $f(\vec{x})$ for any $\alpha>0$. Furthermore, the ${\rm prox}$ operator is necessarily a non-expansive mapping. These properties allow us to use the following iteration formula to find a minimizer of $f(\vec{x})$.
\begin{eqnarray}
\vec{x}^{(k+1)}={\rm prox}_{\alpha f}(\vec{x}^{(k)}),
\end{eqnarray}
{\par\noindent}where $k$ denotes the iteration number. This iterative algorithm is called the proximal minimization algorithm, which provides a powerful framework for non-differentiable convex minimizations.
{\par\noindent}[Proximal Splitting] [5],[6] Let us consider a convex minimization problem formulated as
\begin{eqnarray}
\min_{\vec{x}} f(\vec{x})\equiv\sum_{i=1}^I f_i(\vec{x}),
\end{eqnarray}
{\par\noindent}where $f_i(\vec{x})(i=1,2,\dots,I)$ are possibly non-differentiable lsc component functions and $I\geq 2$. We consider the situation where the ${\rm prox}$ operator corresponding to $f(\vec{x})$ is intensive to compute, but the ${\rm prox}$ operator corresponding to every $f_i(\vec{x})$ can be computed in low costs. The proximal splitting is a framework to construct an iterative algorithm to minimize $f(\vec{x})$ under such situations. In most literatures on the proximal splitting, the number of component functions (splitting) $I$ is 2 [5]. In this paper, we employ a multi-splitting version, {\it i.e.} $I\geq 3$, described and investigated by Passty, because it naturally leads to a row-action-type iterative algorithm which resembles the structure of ART algorithm [6]. Its principle is to compute the prox operator corresponding to each component function $f_i(\vec{x})$ sequentially according to the order $i=1,2,\dots,I$ to obtain the next iterate $\vec{x}^{(k+1)}$ from the previous iterate $\vec{x}^{(k)}$. The iteration formula of Passty proximal splitting is summarized in Algorithm 1.
\begin{table}[h]
\begin{center}
\begin{tabular}{l}
\hline
{\par\noindent}Algorithm 1: Passty Proximal Splitting Algorithm \\
\hline\hline
{\par\noindent}Give an initial vector $\vec{x}^{(0,1)}$. Execute the following. \\
{\par\noindent}loop $\ k=0,1,2,\dots$ \\
{\par\noindent}$\ \ \ $loop $\ i=1,2,\dots,I$ \\
{\par\noindent}$\ \ \ $$\vec{x}^{(k,i+1)}={\rm prox}_{\alpha^{(k)} f_i}(\vec{x}^{(k,i)})$ \\
{\par\noindent}$\vec{x}^{(k+1,1)}=\vec{x}^{(k,I+1)}$ \\
\hline
\end{tabular}
\end{center}
\end{table}
{\par\noindent}In Algorithm 1, $k$ denotes the main iteration number, $i$ denotes the sub iteration number, and $\alpha^{(k)}>0$ is the stepsize parameter dependent on $k$. With respect to the convergence of Algorithm 1, Passty proved the following theorem [6].
{\par\noindent}[Theorem] We consider an ergodic average of the iterates $\vec{x}^{(k',I+1)}(k'=0,1,2,\dots,k)$ in Algorithm 1 defined by
\begin{eqnarray}
\overline{\vec{x}}^{(k)}={\displaystyle{\sum_{k'=0}^k \alpha^{(k')}\vec{x}^{(k',I+1)}}\over{\displaystyle\sum_{k'=0}^k \alpha^{(k')}}}.
\end{eqnarray}
{\par\noindent}Then, $\overline{\vec{x}}^{(k)}$ converges to a minimizer of $f(\vec{x})$ when the diminishing stepsize rule satisfying the following equations is used.
\begin{eqnarray}
\alpha^{(k)}\rightarrow 0\ (k\rightarrow\infty),\ \sum_{k=0}^{\infty}\alpha^{(k)}=\infty,\ \sum_{k=0}^{\infty}(\alpha^{(k)})^2<\infty
\end{eqnarray}
{\par\noindent}We note that the above convergence is called the ergodic convergence, which is weaker than the convergence of sequence $\vec{x}^{(k,I+1)}$ itself to a minimizer of $f(\vec{x})$. In practice, however, we have never observed a situation in which the ergodic convergence occurs but the sequence $\vec{x}^{(k,I+1)}$ itself is not convergent. Therefore, we conjecture that the proposed algorithm can be used without being anxious about the non-convergence.
\par The proposed row-action-type iterative algorithm for the $L^1$-norm reconstruction is derived from Algorithm 1 according to the following steps. First, Eq. (5) is split into the sum of $I$ component cost functions as
\begin{eqnarray}
f(\vec{x})\equiv\sum_{i=1}^I f_i(\vec{x}),\ f_i(\vec{x})=\mid \vec{a}_i^T\vec{x}-b_i \mid (i=1,2,\dots, I),
\end{eqnarray}
{\par\noindent}where $\vec{a}_i$ is the column vector corresponding to the $i$-th row of system matrix $A$. Next, the prox operator corresponding to each $f_i(\vec{x})$ appearing in Algorithm 1 is calculated, which amounts to solving the following optimization problem.
\begin{eqnarray}
\vec{x}^{(k,i+1)}={\rm prox}_{\alpha^{(k)}f_i}(\vec{x}^{(k,i)})\equiv{\rm arg}\min_{\vec{x}}(\mid \vec{a}_i^T\vec{x}-b_i \mid+{{1}\over{2\alpha^{(k)}}}\parallel\vec{x}-\vec{x}^{(k,i)}\parallel_2^2)
\end{eqnarray}
{\par\noindent}This problem can be solved in closed form by using the standard Lagrange multiplier technique [18],[19]. First, by introducing an additional variable $z$, the above minimization problem can be converted into the constrained minimization
\begin{eqnarray}
{\rm arg}\min_{(\vec{x},z)} (\mid z-b_i\mid+{{1}\over{2\alpha^{(k)}}}\parallel\vec{x}-\vec{x}^{(k,i)}\parallel_2^2)\ \ {\rm subject}\ {\rm to}\ \ \vec{a}_i^T\vec{x}=z.
\end{eqnarray}
{\par\noindent}
{\par\noindent}The Lagrangian function corresponding to Eq. (15) is defined by
\begin{eqnarray}
L(\vec{x},z,\lambda)=\mid z-b_i\mid+{{1}\over{2\alpha^{(k)}}}\parallel\vec{x}-\vec{x}^{(k,i)}\parallel_2^2+\lambda(\vec{a}_i^T\vec{x}-z),
\end{eqnarray}
{\par\noindent}where $\lambda$ is the Lagrange multiplier, which is also called the dual variable. The dual problem corresponding to Eq. (16) is defined by
\begin{eqnarray}
\max_{\lambda} D(\lambda)&\equiv&\min_{(\vec{x},z)} L(\vec{x},z,\lambda)\ \ {\rm subject}\ {\rm to}\ \ \lambda\in\Omega \nonumber \\
\Omega&=&\{\lambda\mid D(\lambda)>-\infty\},
\end{eqnarray}
{\par\noindent}where $D(\lambda)$ is the dual function and $\Omega$ is the domain on which $D(\lambda)$ is defined. In the current case, $D(\lambda)$ and $\Omega$ can be explicitly calculated as
\begin{eqnarray}
D(\lambda)&=&\min_{\vec{x}}\min_z[\mid z-b_i\mid+{{1}\over{2\alpha^{(k)}}}\parallel\vec{x}-\vec{x}^{(k,i)}\parallel_2^2+\lambda(\vec{a}_i^T\vec{x}-z)] \nonumber \\
&=&\min_{\vec{x}}(-\lambda b_i+{{1}\over{2\alpha^{(k)}}}\parallel\vec{x}-\vec{x}^{(k,i)}\parallel_2^2+\lambda\vec{a}_i^T\vec{x}) \nonumber \\
&=&-\lambda b_i-{{1}\over{2\alpha^{(k)}}}\parallel\vec{x}^{(k,i)}-\lambda\alpha^{(k)}\vec{a}_i\parallel_2^2+{{1}\over{2\alpha^{(k)}}}\parallel\vec{x}^{(k,i)}\parallel_2^2 \nonumber \\
\Omega&=&\{\lambda\mid -1\leq\lambda\leq 1\},
\end{eqnarray}
{\par\noindent}where the relation between the primal variable $\vec{x}$ and the dual variable $\lambda$ is expressed as
\begin{eqnarray}
\vec{x}=\vec{x}^{(k,i)}-\lambda\alpha^{(k)}\vec{a}_i.
\end{eqnarray}
{\par\noindent}Some additional explanations on the above mathematical calculation are in order. To derive the second equation from the first equation in Eq. (18), we used the fact that $\min_z (\mid z-b_i\mid-\lambda z)=-\infty$ when $\lambda<-1$ or $\lambda>1$ and $\min_z (\mid z-b_i\mid-\lambda z)=-\lambda b_i$ when $-1\leq\lambda\leq 1$ so that the domain of $D(\lambda)$ becomes $\Omega=\{\lambda\mid -1\leq\lambda\leq 1\}$. Furthermore, Eq. (19) is obtained from the minimization with respect to $\vec{x}$ in the second equation in Eq. (18). Therefore, by maximizing $D(\lambda)$ subject to the constraint $\lambda\in\Omega$, the solution to dual problem (Eq. (17)) is given by
\begin{eqnarray}
\lambda=\cases{-1 & (if $q\leq -1$) \cr q & (if $-1<q<1$) \cr 1 & (if $q\geq 1$)},\ 
q=-{{b_i-\vec{a}_i^T\vec{x}^{(k,i)}}\over{\alpha^{(k)}\parallel\vec{a}_i\parallel_2^2}}.
\end{eqnarray}
{\par\noindent}By substituting Eq. (20) into Eq.(19) yields the closed form expression of $\vec{x}^{(k,i+1)}={\rm prox}_{\alpha^{(k)}f_i}(\vec{x}^{(k,i)})$.
\par With respect to the selection of stepsize parameter $\alpha^{(k)}$, we use the diminishing stepsize rule satisfying Eq. (12) expressed as
\begin{eqnarray}
\alpha^{(k)}={{\alpha_0}\over{1+\epsilon k}},
\end{eqnarray}
{\par\noindent}where $(\alpha_0>0,\epsilon>0)$ are pre-specified parameters in the stepsize control. Using the obtained expression of prox operator in the Passty proximal splitting, the proposed algorithm can be summarized in Algorithm 2.
\begin{table}
\begin{center}
\begin{tabular}{p{40em}}
\hline
{\par\noindent}Algorithm 2: $L^1$-Norm Reconstruction \\
\hline\hline
{\par\noindent}Set the stepsize control parameters $(\alpha_0>0,\epsilon>0)$. Give an initial vector $\vec{x}^{(0,1)}$. Execute the following steps for $k=0,1,2,\dots$. \\
{\par\noindent}[Step 1]Set the stepsize parameter $\alpha^{(k)}$
{\par\noindent}$\ \ \displaystyle\alpha^{(k)}={{\alpha_0}\over{1+\epsilon k}}$ \\
{\par\noindent}[Step 2]Execute the following steps for $i=1,2,\dots,I$. \\
{\par\noindent}$\ \ \ (2.1)\ {\rm Compute}\ \lambda$ \\
{\par\noindent}$\ \ \ \ \ \ \displaystyle q=-{{b_i-\vec{a}_i^T\vec{x}^{(k,i)}}\over{\alpha^{(k)}\parallel\vec{a}_i\parallel_2^2}}$ \\
{\par\noindent}$\ \ \ \ \ \ \lambda=\cases{-1 & (if $q\leq -1$) \cr q & (if $-1<q<1$) \cr 1 & (if $q\geq 1$)}$ \\
{\par\noindent}$\ \ \ (2.2)\ {\rm Update}\ \vec{x}$ \\
{\par\noindent}$\ \ \ \ \ \ \vec{x}^{(k,i+1)}=\vec{x}^{(k,i)}-\lambda\alpha^{(k)}\vec{a}_i$ \\
{\par\noindent}[Step 3]$\vec{x}^{(k+1,1)}=\vec{x}^{(k,I+1)}$ \\
\hline
\end{tabular}
\end{center}
\end{table}
\par Algorithm 2 possesses a row-action-type structure similar to the ART algorithm, in which each image update is performed along the direction of $i$-th row of the system matrix $\vec{a}_i$ [20]. In tomographic reconstruction fields, there is a strong trend that people prefer such iterative algorithms, because they can be implemented easily thanks to the necessity of accessing only a single row of $A$ for each image update. Furthermore, it is known that such algorithms can be accelerated by using the special data access order described by Herman and Meyer [8]. We use this data access order in our implementations.
\subsection{Comparison with $L^2$-Norm Reconstruction}
\par The above algorithm derivation can be also used to develop an analogous row-action-type algorithm for the $L^2$-norm reconstruction in Eq. (6). A comparison between the iteration formula for the $L^1$-norm reconstruction and that for the $L^2$-norm reconstruction provides a useful insight together with a nice intuitive interpretation into the proposed algorithm. In this section, we perform such a comparison. First, by using the same algorithm derivation outlined above to the $L^2$-norm cost function in Eq. (6), we obtain the following iteration formula.
\begin{eqnarray}
\lambda&=&-{{2(b_i-\vec{a}_i^T\vec{x}^{(k,i)})}\over{1+2\alpha^{(k)}\parallel\vec{a}_i\parallel_2^2}} \nonumber \\
\vec{x}^{(k,i+1)}&=&\vec{x}^{(k,i)}-\lambda\alpha^{(k)}\vec{a}_i
\end{eqnarray}
{\par\noindent}The iteration formulae for both the $L^1$-norm and $L^2$-norm reconstructions possess a row-action-type structure [20]. The major difference lies in the way to compute the value of variable $\lambda$ which can be interpreted as the amount of update along the vector $\vec{a}_i$. From the description of Algorithm 2 in the $L^1$-norm reconstruction, the value $\lambda$ is truncated to -1 or 1 when the residual error $r_i=\mid b_i-\vec{a}_i^T\vec{x}^{(k,i)}\mid$ is large. On the other hand, from Eq. (22), such truncation is not performed in the $L^2$-norm reconstruction at all. Intuitively, this truncation can be considered a judgement whether the measured data $b_i$ contains an abnormal error or not. When $r_i$ is large, it can be expected that $b_i$ is contaminated by the abnormal error, so the amount of update $\lambda$ is truncated to a small value to avoid overfitting to the erroneous data. On the other hand, when $r_i$ is small, it is expected that $b_i$ is free from the abnormal error, so the amount of update is used without the truncation. Since the $L^2$-norm reconstruction is based on the independent Gaussian noise model, such truncation operation is not necessary because every projection data value is contaminated with noise having the same standard deviation. In conclusion, from the above discussion, we expect that the $L^1$-norm reconstruction is robust against the existence of abnormal data in the projection data. We demonstrate that the difference between the $L^1$-norm reconstruction and the $L^2$-norm reconstruction is surprisingly large in Section 3.
\begin{figure}
\centerline{\includegraphics[height=7cm,clip]{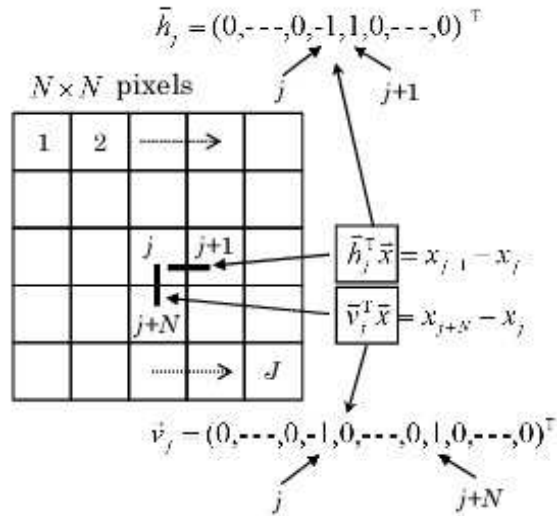}}
\caption{Definitions of the horizontal difference $h_j^T\vec{x}$ and the vertical difference  $v_j^T\vec{x}$ used in the TV norm.}
\end{figure}
\subsection{Improved Algorithm Using Weak Total Variation Penalty}
\par The power to remove artifacts in the proposed algorithm (Algorithm 2) can be strengthened by adding a weak Total Variation (TV) penalty into the cost function $f(\vec{x})$ [9]-[11]. The reason is that the TV penalty imposed on $\vec{x}$ helps in eliminating streak artifacts generated by the abnormal data in image space, in addition to identifying and excluding the abnormal data in projection space. We call this modified algorithm the $L^1$-TV reconstruction, in which the cost function $f(\vec{x})$ is expressed as
\begin{eqnarray}
f(\vec{x})=\beta \parallel\vec{x}\parallel_{\rm TV} + \parallel A\vec{x}-\vec{b} \parallel_1^1\ \  (L^1{\rm -}{\rm TV}\ {\rm reconstruction}),
\end{eqnarray}
{\par\noindent}where $\beta>0$ denotes the hyper parameter to control the strength of TV penalty and $\parallel\vec{x}\parallel_{\rm TV}$ is the TV norm defined by
\begin{eqnarray}
\parallel\vec{x}\parallel_{\rm TV}\equiv\sum_{j=1}^J\sqrt{(\vec{h}_j^T\vec{x})^2+(\vec{v}_j^T\vec{x})^2}.
\end{eqnarray}
{\par\noindent}In Eq. (24), $\vec{h}_j^T\vec{x}$ and $\vec{v}_j^T\vec{x}$ are inner product representations of finite difference operations around the $j$-th pixel along the horizontal and vertical directions, respectively. See Fig. 2 for the detailed definitions of $\vec{h}_j$ and $\vec{v}_j$. The iterative algorithm to minimize Eq. (23) can be derived based on the proximal splitting as follows. First, we split the cost function $f(\vec{x})$ into the sum of $I+1$ component cost functions as
\begin{eqnarray}
f(\vec{x})\equiv\sum_{i=1}^{I+1} f_i(\vec{x}),\ f_i(\vec{x})=\mid \vec{a}_i^T\vec{x}-b_i \mid (i=1,2,\dots, I),\ f_{I+1}(\vec{x})=\beta \parallel\vec{x}\parallel_{\rm TV}.
\end{eqnarray}
{\par\noindent}Applying the Passty proximal splitting to the split in Eq. (25) leads to the iterative algorithm summarized in Algorithm 3, which is similar to Algorithm 2.
\begin{table}
\begin{center}
\begin{tabular}{p{40em}}
\hline
{\par\noindent}Algorithm 3: $L^1$-TV Reconstruction \\
\hline\hline
{\par\noindent}Set the hyper parameter $\beta>0$. Give an initial vector $\vec{x}^{(0,1)}$. Set the stepsize control parameters $(\alpha_0>0,\epsilon>0)$. Execute the following steps for $k=0,1,2,\dots$. \\
{\par\noindent}[Step 1]Set the stepsize parameter $\alpha^{(k)}$
{\par\noindent}$\ \ \displaystyle\alpha^{(k)}={{\alpha_0}\over{1+\epsilon k}}$ \\
{\par\noindent}[Step 2]Execute the following steps for $i=1,2,\dots,I$. \\
{\par\noindent}$\ \ \ (2.1)\ {\rm Compute}\ \lambda$ \\
{\par\noindent}$\ \ \ \ \ \ \displaystyle q=-{{b_i-\vec{a}_i^T\vec{x}^{(k,i)}}\over{\alpha^{(k)}\parallel\vec{a}_i\parallel_2^2}}$ \\
{\par\noindent}$\ \ \ \ \ \ \lambda=\cases{-1 & (if $q\leq -1$) \cr q & (if $-1<q<1$) \cr 1 & (if $q\geq 1$)}$ \\
{\par\noindent}$\ \ \ (2.2)\ {\rm Update}\ \vec{x}$ \\
{\par\noindent}$\ \ \ \ \ \ \vec{x}^{(k,i+1)}=\vec{x}^{(k,i)}-\lambda\alpha^{(k)}\vec{a}_i$ \\
{\par\noindent}[Step 3]Compute the proximity operator for the TV term  \\
{\par\noindent}\ \ \ $\displaystyle \vec{x}^{(k,I+2)}={\rm prox}_{\alpha^{(k)}\beta\parallel\vec{x}\parallel_{\rm TV}}(\vec{x}^{(k,I+1)})={\rm arg}\min_{\vec{x}}(\beta \parallel\vec{x}\parallel_{\rm TV}+{{1}\over{2\alpha^{(k)}}}\parallel\vec{x}-\vec{x}^{(k,I+1)}\parallel_2^2)$ \\
{\par\noindent}[Step 4]$\vec{x}^{(k+1,1)}=\vec{x}^{(k,I+2)}$ \\
\hline
\end{tabular}
\end{center}
\end{table}
{\par\noindent}In Algorithm 3, we need to compute the prox operator corresponding to the TV term (Step 3). This minimization problem is same as that appearing in the so-called TV denoising (ROF model), so that we can use a standard algorithm such as Chambolle's projection algorithm [9],[10]. In our implementations, we used Chambolle's projection algorithm.
\section{SIMULATION STUDIES}
\par We performed simulation studies to demonstrate performances of the proposed $L^1$-norm and $L^1$-TV reconstruction algorithms. We considered three typical scenarios summarized below. In all the scenarios, projection data corresponding to erroneous bins were replaced by values selected from some pre-specified range $[A\vec{x}|_i-m_1,A\vec{x}|_i+m_2]$, where $m_1>0,m_2>0$ are the parameters corresponding to the dynamic range to control the magnitude of errors.
{\par\noindent}[Scenario 1](Detector Errors) Two cases were simulated. In the first case, projection data bins corresponding to two detector elements were contaminated with the abnormal errors (Detector Error 1). In the second case, projection data bins corresponding to two adjacent detectors were contaminated with the abnormal errors in two different locations (Detector Error 2).
{\par\noindent}[Scenario 2](X-ray Radiation Errors) Two cases were simulated. In the first case, projection data bins corresponding to randomly selected angles (10 percent) were contaminated with the abnormal errors (Angle Error 1). In the second case, projection data bins corresponding to randomly selected angles (20 percent containing adjacent angles) were contaminated with the abnormal errors (Angle Error 2).
{\par\noindent}[Scenario 3](Random Errors) Two cases were simulated. In the first case, randomly selected projection data bins (20 percent) were contaminated with the abnormal errors (Random Error 1). In the second case, randomly selected projection data bins (30 percent) were contaminated with the abnormal errors (Random Error 2).
{\par\noindent}In all the scenarios, we used a single transaxial slice of chest CT scan image consisting of 320$\times$320 pixels. The simulated projection data was computed with the sampling of 320(angles)$\times$320(radial bins), from which an image consisting of 320$\times$320(pixels) was reconstructed. We have compared the following four different algorithms.
{\par\noindent}[$L^2$-Norm Reconstruction] The $L^2$-norm reconstruction of Eq. (22) was implemented with the iteration number 50.
{\par\noindent}[Projection Space Median Filter] This is an empirical method to remove the effect of abnormal errors. The median filter is applied to degraded projection data to remove the abnormal errors followed by the $L^2$-norm reconstruction. The window size of median filter was empirically determined in such a way that visual quality of reconstructed image is best dependent on each case.
{\par\noindent}[$L^1$-Norm Reconstruction] The $L^1$-norm reconstruction (Algorithm 2) was implemented with the iteration number 50.
{\par\noindent}[$L^1$-TV Reconstruction] The $L^1$-TV reconstruction (Algorithm 3) was implemented with the iteration number 50.
\par We show reconstructed images together with the degraded projection data for all the scenarios in Figs. 3-5. In all the scenarios, the images by the $L^2$-norm reconstruction were severely damaged by the abnormal errors, in which the artifact patterns, {\it i.e.} streaks, random errors, etc., depend on the locations of abnormal bins. The empirical projection space median filtering succeeded in reducing the artifacts, but the filtering also affects the correct data so that we can observe some additional artifacts in the final images. On the other hand, for the cases of Detector Error 1, Angle Error 1, and Random Error 1 with relatively mild errors, the power of identifying the abnormal bins in both the $L^1$-norm and $L^1$-TV reconstructions were significant in which they succeeded in reconstructing almost perfect images. However, the difference between the $L^1$-norm reconstruction and the $L^1$-TV reconstruction became apparent for the more difficult cases of Detector Error 2, Angle Error 2, and Random Error 2. In these cases, the $L^1$-TV reconstruction correctly identified most of the abnormal bins whereas the $L^1$-norm reconstruction did not succeed perfectly. With respect to convergence speed of the $L^1$-norm and $L^1$-TV reconstructions, they were a bit slower compared with the $L^2$-norm reconstruction and the standard ART algorithm mainly because early iterations need to be spent to correctly identify the locations of abnormal bins. However, thanks to their row-action structures, they seem to be still significantly faster than using the standard $L^1$-norm and $L^1$-TV minimization algorithms (for example, popular iterative reweighted least-squares method [3],[4]) which can be used for the same cost functions.
\begin{figure}
\centerline{\includegraphics[height=12cm,clip]{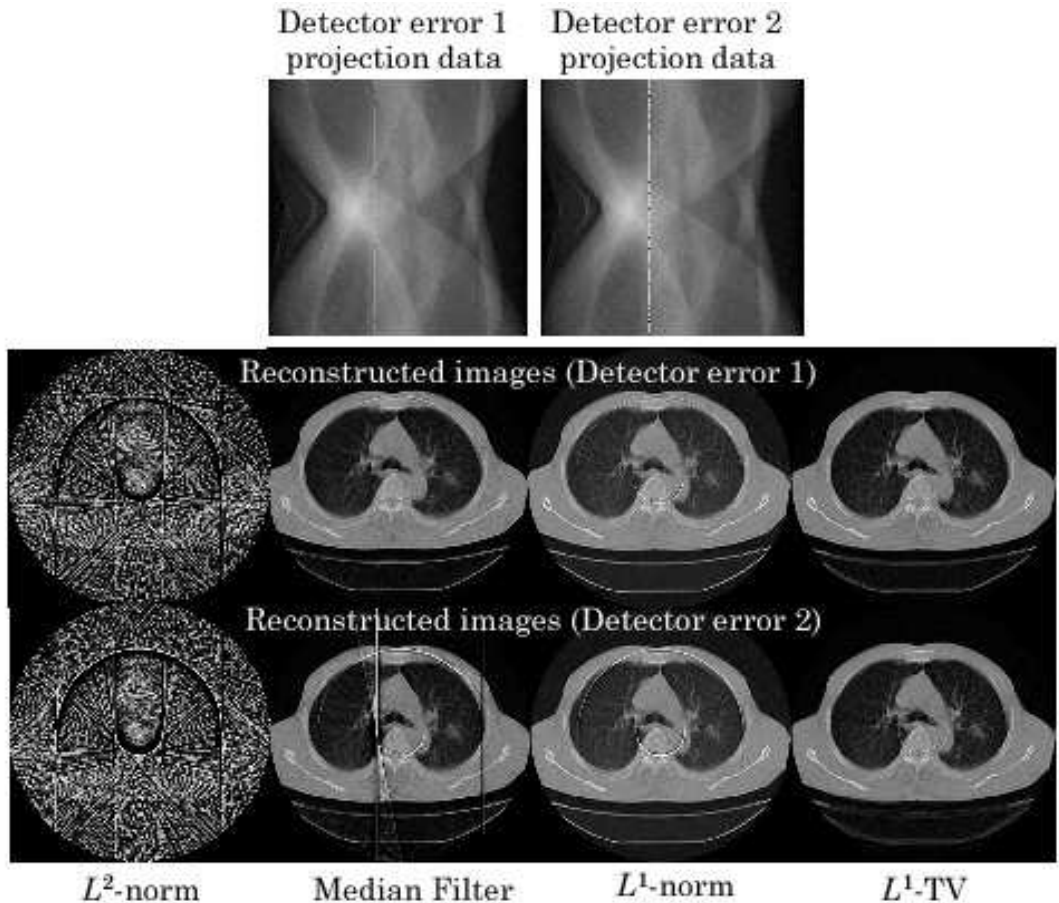}}
\caption{Projection data and corresponding reconstructed images for the scenarios of Detector Error 1 and Detector Error 2.}
\end{figure}
\begin{figure}
\centerline{\includegraphics[height=12cm,clip]{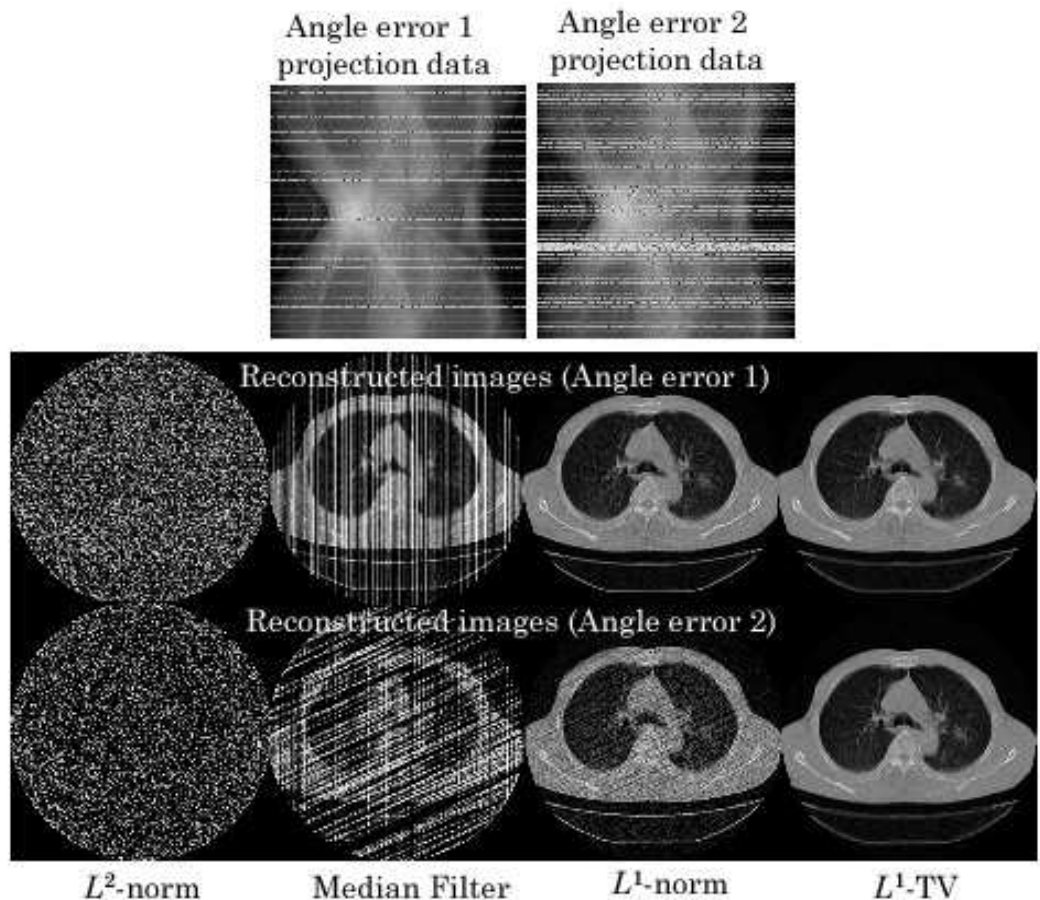}}
\caption{Projection data and reconstructed images for the scenarios of Angle Error 1 and Angle Error 2.}
\end{figure}
\begin{figure}
\centerline{\includegraphics[height=12cm,clip]{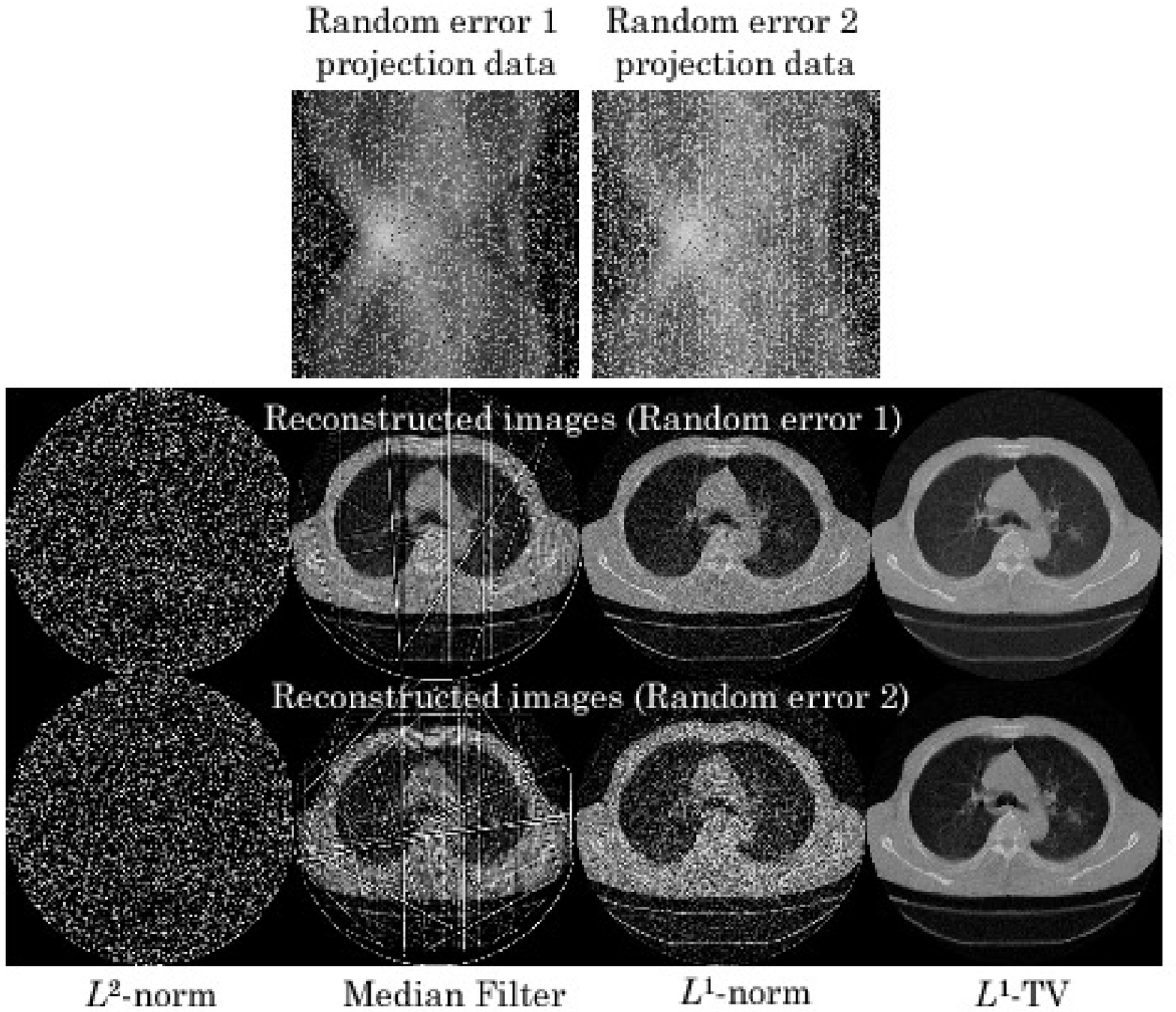}}
\caption{Projection data and reconstructed images for the scenarios of Random Error 1 and Random Error 2.}
\end{figure}
\section{CONCLUSIONS}
\par In this paper, we proposed a new image reconstruction algorithm called the Fault-Tolerant reconstruction for tomographic imaging situation where a small number of projection data bins are contaminated by abnormal errors occurred during the measurement by some physical reasons. More specifically, we derived two row-action-type iterative algorithms called the $L^1$-norm and $L^1$-TV reconstructions. In the simulation studies, compared with the naive approaches such as the $L^2$-norm reconstruction and the projection space median filter (followed by the $L^2$-norm reconstruction), we demonstrated that the proposed algorithms are robust to the existence of abnormal projection data bins. We believe that this work is valuable because there exist very few (almost no, to the best of our knowledge) works to handle tomographic reconstruction under the existence of abnormal errors.
\acknowledgments
This work was partially supported by JSPS KAKENHI Grant Number 15K06103.

\end{document}